\documentclass[Physsubmission, Phys]{SciPost}

\hypersetup{
    colorlinks,
    linkcolor={red!50!black},
    citecolor={blue!50!black},
    urlcolor={blue!80!black}
}
\usepackage{mathtools}
\usepackage{amsmath}
\usepackage{amssymb}
\usepackage{physics}
\usepackage[bitstream-charter]{mathdesign}
\DeclareSymbolFont{usualmathcal}{OMS}{cmsy}{c}{n}
\DeclareSymbolFontAlphabet{\mathcal}{usualmathcal}
\renewcommand{\v}[1]{ \ensuremath{ {\underline{#1}} }}

\begin{document}

\begin{center}{\Large \textbf{
Entanglement, partial set of measurements, and diagonality of the density matrix in the parton model  \\
}}\end{center}

\begin{center}
Haowu Duan\textsuperscript{1},
\end{center}

\begin{center}
{\bf 1} North Carolina State University
* hduan2@ncsu.edu
\end{center}

\begin{center}
\today
\end{center}


\definecolor{palegray}{gray}{0.95}
\begin{center}
\colorbox{palegray}{
  \begin{tabular}{rr}
  \begin{minipage}{0.1\textwidth}
    \includegraphics[width=22mm]{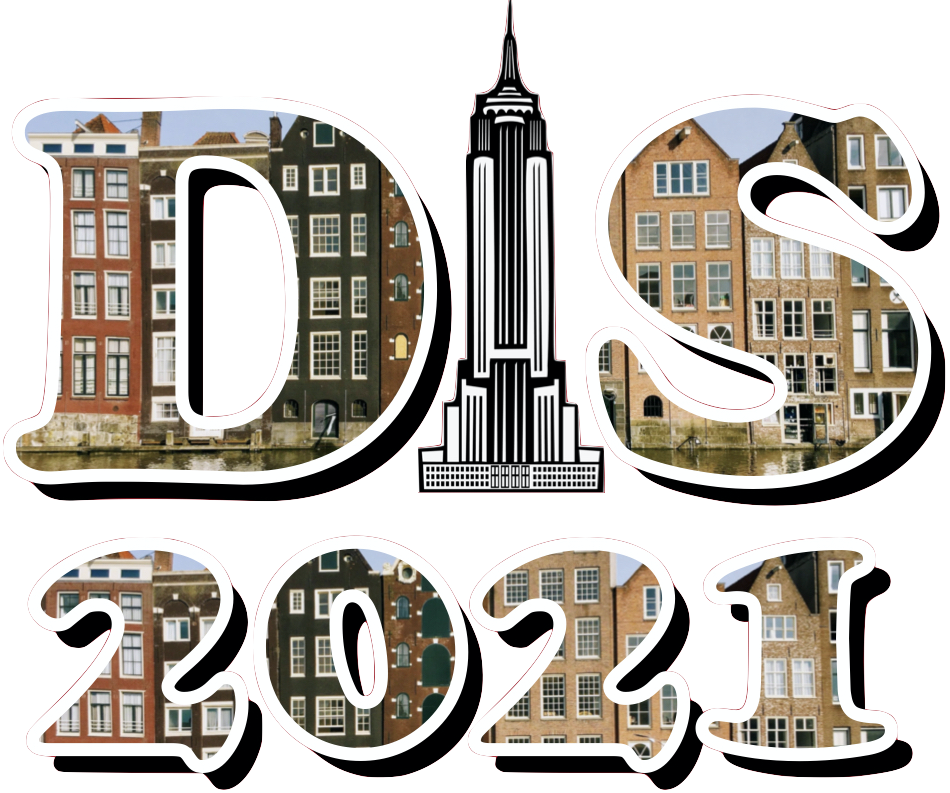}
  \end{minipage}
  &
  \begin{minipage}{0.75\textwidth}
    \begin{center}
    {\it Proceedings for the XXVIII International Workshop\\ on Deep-Inelastic Scattering and
Related Subjects,}\\
    {\it Stony Brook University, New York, USA, 12-16 April 2021} \\
    \doi{10.21468/SciPostPhysProc.?}\\
    \end{center}
  \end{minipage}
\end{tabular}
}
\end{center}

\section*{Abstract}
{\bf
To study quantum properties of the  hadron wavefunction at small x, we derived  the reduced density matrix for soft gluons in the CGC framework. We explicitly showed that the reduced density matrix is not diagonal in the particle number basis. The off-diagonal components are usually ignored in the conventional parton model. We thus defined the density matrix of ignorance 
by keeping only the part of the reduced density matrix which can be probed in a limited set of experimental measurements. We calculated Von Neumann entropy for both the reduced and ignorance density matrices. The entropy of ignorance is always greater than  the entanglement entropy (computed drom the reduced density matrix) of gluons. Finally, we showed that the CGC reduced density matrix for soft gluons can be diagonalized in a thermal basis with Boltzmann weights suggesting thermalization of new quasi-particle states which we  dubbed entangolons.  
}


\section{Introduction} 

 This work was motivated by the paradox raised in  Ref.~\cite{Kharzeev:2017qzs}. On one hand the proton as a quantum object is in a pure state and is described by a completely coherent wavefunction with zero entropy. On the other hand in high energy experiments when probed by a small external probe, it behaves like an incoherent ensemble of (quasi-free) partons. Such an ensemble carries a nonvanishing entropy. Reference~\cite{Kharzeev:2017qzs} suggested that the origin of this entropy is entanglement between the degrees of freedom one observes in DIS (partons in the small spatial region of the proton) and the rest of the proton wavefunction which are not measured in the final state and therefore play the role of  an ``environment''. 
 Formally, this can be formulated  in terms of the reduced density matrix 
\begin{equation}
\hat\rho_{r}={\rm Tr}_{\rm unobs}\, \Big[\, |P\rangle\langle P| \, \Big]\,,
\end{equation}
where $|P\rangle$ is the proton wave function and the partial trace is taken over the unobserved degrees of freedom. The entropy of the parton model is then identified with the von Neumann entropy of the reduced density matrix according to
\begin{equation}
S_{\rm PM}=-{\rm Tr} \Big[\, \hat\rho_{r}\, \ln \hat\rho_{ r}\, \Big]\,.
\end{equation}
Most of the quantities measured DIS are related to  the average number of particles or 
 multi-parton momentum distributions  $ \langle a^\dagger(\v{k}_1)a(\v{k}_1)...a^\dagger(\v{k}_n)a(\v{k}_n) \rangle $. These quantities  do not provide an  access to  the off-diagonal elements of the density matrix  in the number operator basis. 
 This leads to the fundamental problem that  one cannot reconstruct the full density matrix from the available experimental data.

This lack of knowledge of the actual density matrix of the system can be characterized by an entropy. We will dub this entropy ``{\it the entropy of ignorance}''. Among the family of all density matrices that reproduce the same experimental observable, we defined the matrix that have zeros on all off-diagonal elements as the density matrix of ignorance (the operation of dropping the off-diagonal elements of the density matrix can be considered as an example of a quantum-to-classical channel). The associated entropy 
 \begin{equation}
 S_I=-{\rm Tr}\Big[\hat\rho_I\ln\hat\rho_I\Big]\,.
 \end{equation}
 It has the following property $S_I \ge  -{\rm Tr}\Big[\hat\rho_r\ln\hat\rho_r\Big]$ (see e.g. Refs.~\cite{Wehrl:1978zz,Witten:2018zva}). The entropy of ignorance is exactly equal to the Boltzmann entropy of the classical ensemble of partons with the probability distribution defined by the  diagonal matrix elements 
\begin{equation}
S_I=S_B=-\sum_n\, p_n\ln p_n; \quad \quad p_n= \langle n|\hat\rho_r|n \rangle \,.
\end{equation}

\section{The CGC density matrices }
The Color Glass Condensate (CGC) effective theory describes scattering at high energy  (see Refs.~\cite{Iancu:2002xk,Kovner:2005pe,McLerran:2008uj,Gelis:2010nm,Kovchegov:2012mbw} for reviews).
According to the CGC, the valence (``hard'') partons can be treated as static classical sources of the small-x ``soft'' gluons. This partial separation of degrees of freedom can be formalized by  
\begin{equation}
| \psi \rangle = | s \rangle  \otimes   | v \rangle  \,,
\end{equation}
where $| v \rangle $ is the  state vector characterizing the valence degrees of freedom and  $| s \rangle$ is the soft fields in the presence of the valence source. Despite appearances, the state is not of a direct product form since the soft vacuum depends on the valence degrees of freedom. In the leading perturbative order, the CGC soft gluon state is
\begin{equation}
    | s \rangle=\mathcal{C}| 0\rangle
\end{equation}
with the coherent operator
\begin{equation}
	\label{Eq:CO}
	{\cal C}=\exp\left\{2 i {\rm tr} \int \frac{d^2k }{(2\pi)^2} b^i(\v{k}) \phi_i(\v{k}) \right\}=\exp\left\{2 i {\rm tr} \int_{\v{k}} b^i(\v{k}) \phi_i(\v{k}) \right\}\,,
\end{equation}
where $\phi_i(\v{k})\equiv a_i^+(\v{k})+a_i(-\v{k})$, and the trace operation (tr)  is  over all colors. 
The background field $b^i_a$ is determined by the valence color charge density $\rho$ via:
\begin{equation}
    b^i_a(\v{k})=g\rho_a(\v{k})\frac{i\v{k}_i}{k^2} + {\cal O}(\rho^2)\,.
	\label{Eq:b}
\end{equation}
For simplicity we restrict our consideration only to the leading order contribution in the charge density  $\rho$ which captures only the longitudinal  polarization of the soft gluons. We do not expect our results to change if higher order corrections are included. 

The valence wave function $|v\rangle$ is customarily modeled following the  McLerran-Venugopalan (MV) model~\cite{McLerran:1993ni,McLerran:1993ka}
\begin{equation}\label{mv}
  \langle \rho | v \rangle \langle v |  \rho \rangle=   {\mathcal N}e^{-\int_{\v{k}}
 \frac{1}{2\mu^2} \rho_a(\v{k})   
  \rho^*_a(\v{k})}\,,
\end{equation}
where ${\mathcal N}$ is the normalization factor and the parameter $\mu^2$ determines the average color charge density.
Now we can formally define  the hadron density matrix:
\begin{equation}
    \hat{\rho}=|v\rangle\otimes  |s\rangle \langle s| \otimes \langle v| \,.
\end{equation}
Integrating out the valence  degrees of freedom, we obtain the reduced density matrix for the soft gluons:
\begin{align}
    \hat{\rho}_r=\mathcal{N}\int D \rho\, \,  e^{-\int_{\v{k}}\frac{1}{2\mu^2}\rho_a(\v{k})\rho^*_a(\v{k})}\mathcal{C}(\rho_b,\phi_b^i)|0\rangle \langle0|\mathcal{C}^\dagger(\rho_c,\phi_c^j) \, . 
\end{align}

To extract the ignorance density matrix, $\hat \rho_I$, in gluon number basis, we need the basis vectors 
\begin{align}
 \prod_{c}\prod_{k}|n_c( \v{k}), m_c(-\v{k})\rangle\propto \prod_{c}\prod_{k} \left(\frac{[a^{\dagger}_c(\v{k})]^{n_c}}{\sqrt{n_c!}} \right) \left(\frac{[a^{\dagger}_c(-\v{k})]^{m_c}}{\sqrt{m_c!}}\right)|0\rangle\,
\end{align}
where $c$ is the  color indices, and we only consider longitudinal polarization. Since a mode with momentum $\v{k}$ mixes only with the mode with momentum $-\v{k}$ due to the fact that  $\rho^*_a(\v{k})=\rho_a(-\v{k})$, we introduce $N_c=n_c+m_c$. Furthermore, $\hat\rho_r$ is a direct product of density matrices in a fixed transverse momentum sector due to transnational in variance, see \cite{Duan:2020jkz} for details.

The matrix elements of the density matrix for a given value of momentum $q$ and given color c are 
\begin{align}\label{matel}
     & \langle n_c(\v{q}), m_c(-\v{q}) |\hat{\rho}_r(\v{q})|\alpha_c(\v{q}), \beta_c(-\v{q})\rangle 
	 = (1-R) 
      \frac{
      \left( n+\beta\right)!
  }{\sqrt{n!m!\alpha!\beta!}}\left( \frac {R}{2} \right)^{n+\beta} 
      \delta_{ \left( n+\beta \right), \left(m+\alpha\right)} \,,\\  &R =\left(1+\frac{q^2}{2g^2 \mu^2} \right)^{-1}\,.
\end{align}
We thus explicitly demonstrated that the off-diagonal components are non zero!  This is one of the main results of this paper.  If we ignore the off-diagonal elements  we obtain the ignorance density matrix in the number of gluons representation 
 \begin{align}
  \langle n_c(\v{q}), m_c(-\v{q}) |\hat{\rho}_I(\v{q})|n_c(\v{q}), m_c(-\v{q})\rangle
 = (1-R)\frac{\left( n+m\right)!
  }{n!m!}\left( \frac {R}{2} \right)^{n+m} \,. 
 \end{align}

\section{Quantum Entropies}
\begin{figure}[ht]
	\centering 
	\includegraphics[width=0.45\textwidth]{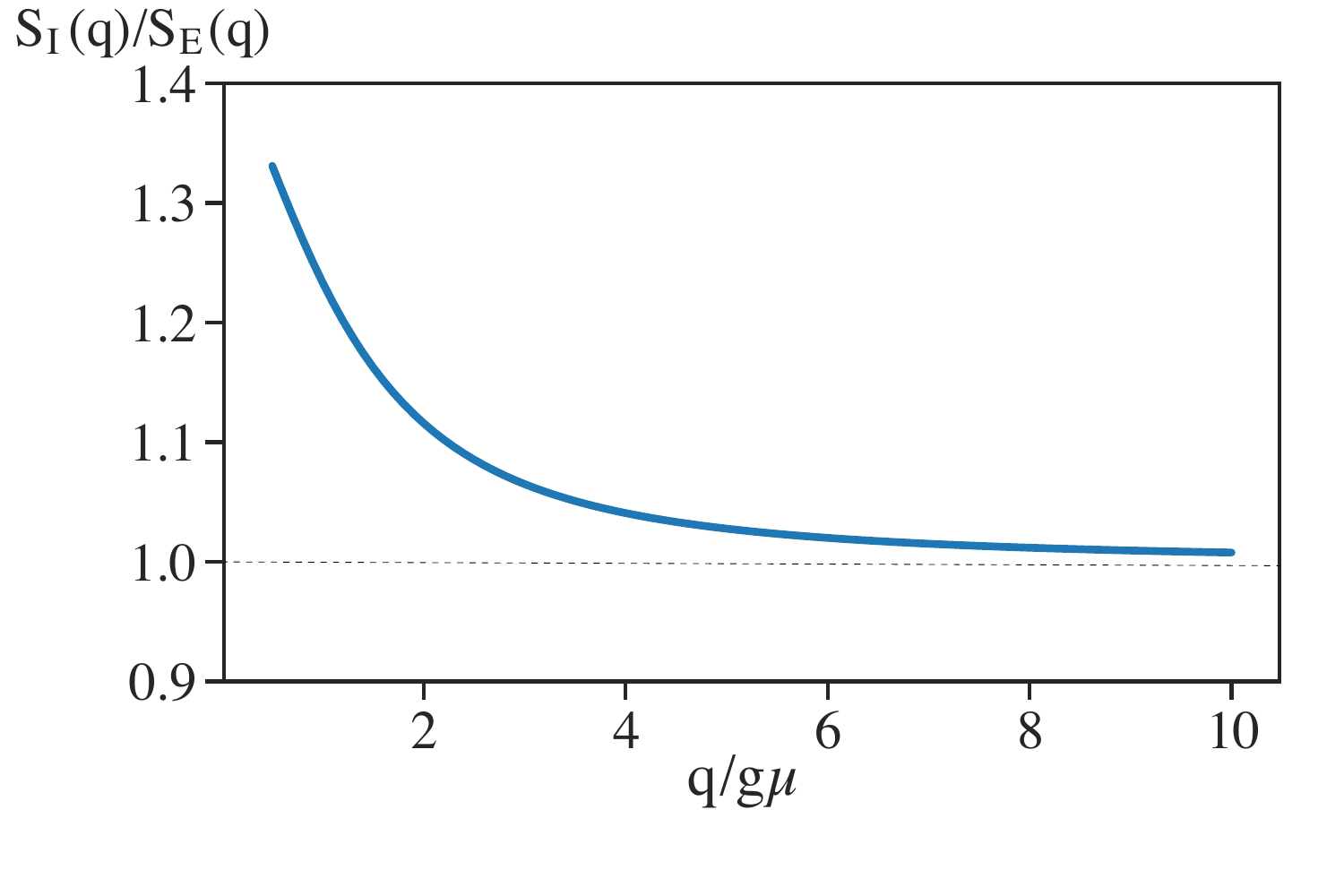}
	\caption{Ratios of entropy densities at a given magnitude of the transverse momentum $q/g\mu$. $S_I(q^2)$ is the von Neumann entropy density of ignorance and  $S_E(q^2)$ is the corresponding entanglement entropy density.}
	\label{fig:Ratio}
\end{figure}

Using the reduced density matrix, the entropy of entanglement can be readily calculated. It reproduces the known result obtained in Ref.~\cite{Kovner:2015hga}:
\begin{align}
        S_E &=\frac{(N_c^2-1) S_\perp}{2} \int \frac{d^2q}{(2\pi)^2} \Bigg[\ln(\frac{g^2\mu^2}{q^2})+ \sqrt{1+4\frac{g^2\mu^2}{q^2}} \notag \\ & \times  \ln (1+\frac{q^2}{2 g^2\mu^2}+ \frac{q^2}{2 g^2\mu^2}\sqrt{1+4\frac{g^2\mu^2}{q^2}} )  \Bigg]\,.
\end{align}
There is no analytic expression for the entropy of ignorance. Hence, we compare the results numerically in Fig.~\ref{fig:Ratio}.
According to the ratio $S_I/S_E$, the dominant contribution to their difference is at small q where states with more than one particle cannot be ignored. While at large q, only the vacuum state survives and the two density matrix are approximately the same hence the ratio approaching 1. 

\section{Thermal quasi-particle basis}
It is amusing that the entanglement entropy can be rewritten in the form 
\begin{align}
    S_E = (N_c^2-1) S_\perp\int \frac{d^2q}{(2\pi)^2} \Bigg[
    (1+f) \ln (1+f) - f \ln f 
    \Bigg]
\end{align}
with  $f = (\exp(\beta \omega) -1 )^{-1}$ and 
$ 
\omega \beta = 2 \ln \left( \frac{ q}{2g \mu}  + \sqrt {  1 + \left( \frac{ q}{2g \mu} \right)^2} \right) 
$. 
This suggests that there exists a quasi-particle basis in the reduced density matrix is diagonal and thermal (i.e. all eigenvalues of the density matrix are powers of the same number.).  We named this quasi-particles entangalons. 

At small momentum, $
\omega \beta  \approx   \frac{ q}{g \mu}$, we thus can identify the effective temperature of entangolons with $g\mu$. 

At large momentum, the quasi-particle distribution function coincides with the Weizsacker-Williams distribution of gluons $f \approx g^2 \mu^2/q^2$.

\section{Conclusions} 
Our explicit calculation confirms that, within the CGC framework, the reduced density matrix for soft gluons is not diagonal in parton/particle number basis. Neglecting the off-diagonal elements results in the loss of information and thus in larger entropy. We introduced a new form of entropy associated with this loss and called it entropy of ignorance.  It is greater than the entanglement entropy which is supposedly related to the entropy of final state particles. In addition, by analysing the entanglement entropy, we hypothesized that   the reduced density matrix directly is diagonal in some quasi-particle basis with the eigenvalues having the form of Boltzmann weights. 

\section{Acknowledgements}
We thanks A. Kovner and V. Skokov for collaborating on this project. 
This works is supported  by the DOE Office of Nuclear Physics through Grant No. DE-SC0020081.

\bibliographystyle{SciPost_bibstyle}

\begin{thebibliography}{10}
\providecommand{\url}[1]{\texttt{#1}}
\providecommand{\urlprefix}{URL }
\expandafter\ifx\csname urlstyle\endcsname\relax
  \providecommand{\doi}[1]{doi:\discretionary{}{}{}#1}\else
  \providecommand{\doi}{doi:\discretionary{}{}{}\begingroup
  \urlstyle{rm}\Url}\fi
\providecommand{\eprint}[2][]{\url{#2}}

\bibitem{Kharzeev:2017qzs}
D.~E. Kharzeev and E.~M. Levin,
\newblock \emph{{Deep inelastic scattering as a probe of entanglement}},
\newblock Phys. Rev. \textbf{D95}(11), 114008 (2017),
\newblock \doi{10.1103/PhysRevD.95.114008},
\newblock \eprint{1702.03489}.

\bibitem{Wehrl:1978zz}
A.~Wehrl,
\newblock \emph{{General properties of entropy}},
\newblock Rev. Mod. Phys. \textbf{50}, 221 (1978),
\newblock \doi{10.1103/RevModPhys.50.221}.

\bibitem{Witten:2018zva}
E.~Witten,
\newblock \emph{{A Mini-Introduction To Information Theory}}  (2018),
\newblock \eprint{1805.11965}.

\bibitem{Iancu:2002xk}
E.~Iancu, A.~Leonidov and L.~McLerran,
\newblock \emph{{The Color glass condensate: An Introduction}},
\newblock In \emph{{QCD perspectives on hot and dense matter. Proceedings, NATO
  Advanced Study Institute, Summer School, Cargese, France, August 6-18,
  2001}}, pp. 73--145 (2002), \eprint{hep-ph/0202270}.

\bibitem{Kovner:2005pe}
A.~Kovner,
\newblock \emph{{High energy evolution: The Wave function point of view}},
\newblock Acta Phys. Polon. \textbf{B36}, 3551 (2005),
\newblock \eprint{hep-ph/0508232}.

\bibitem{McLerran:2008uj}
L.~McLerran,
\newblock \emph{{The Color Glass Condensate and Glasma}}  (2008),
\newblock \eprint{0804.1736}.

\bibitem{Gelis:2010nm}
F.~Gelis, E.~Iancu, J.~Jalilian-Marian and R.~Venugopalan,
\newblock \emph{{The Color Glass Condensate}},
\newblock Ann. Rev. Nucl. Part. Sci. \textbf{60}, 463 (2010),
\newblock \doi{10.1146/annurev.nucl.010909.083629},
\newblock \eprint{1002.0333}.

\bibitem{Kovchegov:2012mbw}
Y.~V. Kovchegov and E.~Levin,
\newblock \emph{{Quantum chromodynamics at high energy}},
\newblock Camb. Monogr. Part. Phys. Nucl. Phys. Cosmol. \textbf{33}, 1 (2012),
\newblock \doi{10.1017/CBO9781139022187}.

\bibitem{McLerran:1993ni}
L.~D. McLerran and R.~Venugopalan,
\newblock \emph{{Computing quark and gluon distribution functions for very
  large nuclei}},
\newblock Phys. Rev. \textbf{D49}, 2233 (1994),
\newblock \doi{10.1103/PhysRevD.49.2233},
\newblock \eprint{hep-ph/9309289}.

\bibitem{McLerran:1993ka}
L.~D. McLerran and R.~Venugopalan,
\newblock \emph{{Gluon distribution functions for very large nuclei at small
  transverse momentum}},
\newblock Phys. Rev. \textbf{D49}, 3352 (1994),
\newblock \doi{10.1103/PhysRevD.49.3352},
\newblock \eprint{hep-ph/9311205}.

\bibitem{Duan:2020jkz}
H.~Duan, C.~Akkaya, A.~Kovner and V.~V. Skokov,
\newblock \emph{{Entanglement, partial set of measurements, and diagonality of
  the density matrix in the parton model}},
\newblock Phys. Rev. D \textbf{101}(3), 036017 (2020),
\newblock \doi{10.1103/PhysRevD.101.036017},
\newblock \eprint{2001.01726}.

\bibitem{Kovner:2015hga}
A.~Kovner and M.~Lublinsky,
\newblock \emph{{Entanglement entropy and entropy production in the Color Glass
  Condensate framework}},
\newblock Phys. Rev. \textbf{D92}(3), 034016 (2015),
\newblock \doi{10.1103/PhysRevD.92.034016},
\newblock \eprint{1506.05394}.

\end{thebibliography}

\nolinenumbers

\end{document}